# Automated surgical planning with nnU-Net: delineation of the anatomy in hepatobiliary phase MRI

*Karin A. Olthof[1,*] Matteo Fusaglia[1], Bianca Güttner[2,3], Tiziano Natali[1], Bram Westerink[4], Stefanie Speidel[2,3], Theo J.M. Ruers[1,5], Koert F.D. Kuhlmann[1], Andrey Zhylka[1]*

[1] Department of Surgical Oncology, The Netherlands Cancer Institute – Antoni van Leeuwenhoek, Amsterdam, The Netherlands

[2] Department of Translational Surgical Oncology, National Center for Tumor Diseases (NCT), NCT/UCC Dresden, a partnership between DKFZ, Faculty of Medicine and University Hospital Carl Gustav Carus, TUD Dresden University of Technology, and Helmholtz-Zentrum Dresden-Rossendorf (HZDR), Dresden, Germany

[3] The Centre for Tactile Internet with Human-in-the-Loop (CeTI), TUD Dresden University of Technology, Dresden, Germany

[4] Department of Radiology, The Netherlands Cancer Institute – Antoni van Leeuwenhoek, Amsterdam, The Netherlands

[5] Faculty of Science and Technology (TNW), Nanobiophysics Group (NBP), University of Twente, Enschede, The Netherlands

**\*** Corresponding author: ka.olthof@nki.nl, +3120 512 2531

**Abstract**

*Background:* 3D models for liver surgery provide a patient-specific spatial understanding of complex hepatic anatomy, enhancing preoperative planning and intraoperative decision-making. The aim of this study was to develop and evaluate a deep learning-based automated segmentation method for hepatic anatomy (i.e., parenchyma, tumors, portal vein, hepatic vein and biliary tree) from the hepatobiliary phase of gadoxetic acid-enhanced MRI. This method should ease the clinical workflow of preoperative planning.

*Methods:* Manual segmentation was performed on hepatobiliary phase MRI scans from 90 consecutive patients who underwent liver surgery between January 2020 and October 2023. A deep learning network (nnU-Net v1) was trained on 72 patients with an extra focus on thin structures and topography preservation. Performance was evaluated on an 18-patient test set by comparing automated and manual segmentations using Dice similarity coefficient (DSC). Following clinical integration, 10 segmentations (assessment dataset) were generated using the network and manually refined for clinical use to quantify required adjustments using DSC.

*Results:* In the test set, DSCs were 0.97±0.01 for liver parenchyma, 0.80±0.04 for hepatic vein, 0.79±0.07 for biliary tree, 0.77±0.17 for tumors, and 0.74±0.06 for portal vein. Average tumor detection rate was 76.6±24.1%, with a median of one false-positive per patient. The assessment dataset showed minor adjustments were required for clinical use of the 3D models, with high DSCs for parenchyma (1.00±0.00), portal vein (0.98±0.01) and hepatic vein (0.95±0.07). Tumor segmentation exhibited greater variability (DSC 0.80±0.27). During prospective clinical use, the model detected three additional tumors initially missed by radiologists.

*Conclusions:* The proposed nnU-Net-based segmentation method enables accurate and automated delineation of hepatic anatomy. This enables 3D planning to be applied efficiently as a standard-of-care for every patient undergoing liver surgery.

*Trial registration:* IRB-d24-254, registration date 8-9-2024

**Background**

The primary aim of liver surgery is to achieve a radical resection or complete ablation, ensuring optimal oncological outcomes. As systemic treatment regimens become more effective, local treatment options are expanding, resulting in a growing emphasis on precision targeting and parenchyma-sparing techniques (1,2). These challenging procedures require advanced preoperative planning tools. Conventional CT or MRI visualizations no longer meet these requirements and patient-specific 3-dimensional (3D) models have been adopted in state-of–the-art hepatobiliary surgery (3–5). These models provide intuitive visualizations of spatial relationships between tumors, vasculature, and other critical structures. Combined with instrument tracking and registration methods, 3D models serve as a map for image-guided procedures (6,7).

Generating 3D models involves manually delineating anatomical structures from diagnostic scans, a process that is labor intensive and reliant on clinical expertise. Deep learning has emerged as an effective solution to automate segmentation. Most models are trained on CT images (8,9), as MRI presents variable signal intensities, susceptibility to artifacts and high heterogeneity in imaging protocols, making segmentation more challenging. Nonetheless, from a clinical perspective, MRI is increasingly preferred over CT because of its superior lesion differentiation and higher sensitivity for detecting small (<1 cm) lesions (10–12). Additionally, it has the capability to image bile ducts, allowing anatomical variations to be detected.

Existing literature on automation of MRI liver segmentation primarily focuses on parenchyma segmentation (13–16) with limited work extending to segmentation of intrahepatic anatomy (17–20). In (17) vasculature was segmented from non-contrast T1 MRI, yet the absence of contrast restricted the ability to segment small peripheral veins, tumors and bile ducts. Alternatively, (18) automated segmentation of the parenchyma, vasculature, bile ducts and tumors from contrast-enhanced MRI. While the network demonstrated good performance for parenchyma segmentation, accuracy for vessels and bile ducts could be improved. In addition, neither of these studies focused on preserving vessel topology (i.e., tree structure). Such an accurate anatomy representation is essential for image-guided procedures, as vessel bifurcations are often used as landmarks for registering the 3D model.

At our institute, complex liver surgery is performed as a standard-of-care. 3D models are used during planning and image-guided surgery to aid the localization of small and vanished lesions. To ease the clinical workflow of preoperative planning, this study aims to develop a more accurate and fast automatic segmentation method of the hepatic anatomy from the hepatobiliary phase of a gadoxetic acid enhanced MRI using deep learning.

**Methods**

*Study population*
Institutional review board approval for this single-center, retrospective study was obtained in September 2024 (IRB-d24-254). This study included patients of 18 years and older undergoing liver surgery between January 2020 and October 2023 in the Netherlands Cancer Institute in Amsterdam. Written informed consent was obtained from all subjects treated within the study protocol for surgical navigation, which was approved by the institutional medical ethics committee in July 2018 (NL65724.031.18). Patients from June 2022 and later were treated using surgical navigation as standard clinical practice at our institute based on indication. Patients were included when they had preoperative 3D models segmented from the hepatobiliary phase of a gadoxetic acid enhanced MRI (Figure 1a). Clinical information was collected from the patients, including sex, age, tumor type, number of lesions, state of the non-affected liver parenchyma (i.e., normal or steatotic) and history of prior hepatic interventions.

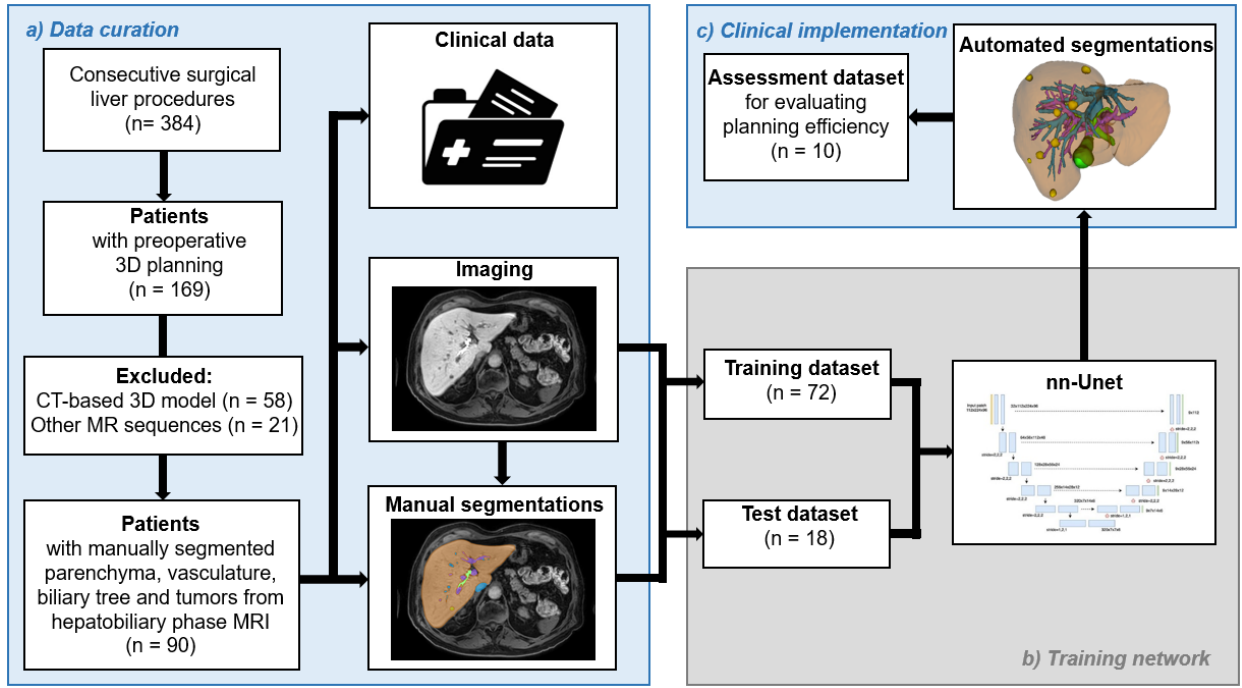

**Figure 1:** Flow diagram depicting the study cohort selection from 169 consecutively performed image-guided liver procedures. After exclusion of CT and non-hepatobiliary phase MR scans, 90 hepatobiliary phase MR images were manually segmented. MRI scans and corresponding segmentations were divided into a training dataset (n = 72) and test dataset (n = 18) to train a nn-Unet. Following clinical integration, automated segmentations (n = 10) generated by the model were used for assessment of planning efficacy.

*MR imaging*
Images used in this study were hepatobiliary phase (20 minutes post-injection) images of a 10 mL Gd-EOB-DTPA (Primovist, Bayer AG, Germany)-enhanced MRI. The scan is acquired using a 3T Philips scanner with a 3D T1-weighted FFE-mDixon sequence, with a 12-second breath-hold scan in expiration, and a voxel size of 2.0x2.0x3.0 mm. Scanning characteristics can be found in Table 1.

**Table 1:** Characteristics of magnetic resonance imaging

| Scanning characteristic | |
|---|---|
| Scanner model | Philips |
| Field strength (T) | 3 |
| Pixel spacing (mm) | -1.5 [1] |
| Slice thickness (mm) | 3 |
| Voxel size (mm) | 2.0x2.0x3.0 [2] |

[1] Negative pixel spacing indicates partially overlapping slices.
[2] Reconstructed voxel size: 1.0x1.0x1.5

*Manual segmentations*
MR imaging was extracted as Digital Imaging and Communications in Medicine (DICOM), anonymized by removing patient identifiers and converted to Neuroimaging Informatics Technology Initiative (NIfTI) format. Segmentation of the hepatobiliary anatomy was performed manually by two technical physicians with over four years of experience in liver segmentation and confirmed by a hepatobiliary surgeon. Tumor segmentation was conducted based on annotations provided by a board-certified radiologist. Segmentation was performed in 3D Slicer(21), an open source medical image analysis software.

*Automated segmentation model*
The dataset consisting of 90 patients was randomly split into training (72 patients) and test (18 patients) sets. The nnUNet (v1) (22) framework was used to train the segmentation model for 500 epochs (Figure 1b). The choice of loss function was determined by the need to achieve high accuracy in delineating thin vein bifurcations as well as maintaining the topology of the vessel tree. Thus, a combination of clDice and bootstrapped cross-entropy was used (23–25).
The clDice loss, developed specifically for vessel segmentation, enforces the preservation of the structure and the connectivity of the vascular tree by computing skeletons of the ground truth and predicted vasculatures and comparing them against the masks: ground truth skeleton against predicted mask and predicted skeleton against ground truth mask.
The use of bootstrapped cross-entropy was inspired by Liew, et al. (23) who used it for thin object delineation. In their definition, bootstrapped cross-entropy essentially focuses on top $K$ voxels with the highest loss value. In our implementation, the use of bootstrapped cross-entropy was preceded by a "warm-up" period of 400 epochs with $K$ set to 100% (regular cross entropy loss). During the following 100 epochs $K$ linearly grew from 15% to 50% in order to avoid overfitting of the model. For each model training run, the weights were randomly initialized.

*Evaluation*
Performance of the model in delineating liver parenchyma, tumor, portal vein, hepatic vein, and bile duct was evaluated against the manual delineations using the Dice similarity coefficient (DSC). The vascular segmentations were categorized into central (i.e., main trunks and primary branches directly arising from them) and peripheral (i.e., all smaller branches originating from central vessels). The biliary tree segmentation was subdivided into bile ducts and the gallbladder. In cholecystectomy cases, central biliary tree segmentation was left out of the analysis.
The network was then integrated into clinical practice to improve the workflow for preoperative modeling (Figure 1c). Segmentations were generated using the network and manually refined to create 3D models for image-guided liver surgery of 10 patients (i.e., the assessment dataset). The automated segmentations were then compared to the manually adjusted models using the DSC to assess the efficiency of assisted preoperative planning.

*Statistical analysis*
Statistical analysis of the DSC scores was performed using the Python ecosystem (Python 3.6.12, SciPy 1.5.4) using unpaired Mann-Whitney test. In addition, patient demographics were compared across the training, test and assessment datasets to ensure group comparability, with statistical analysis conducted in SPSS v25.0® (IBM Corporation; Armonk, NY, USA). Categorical variables were compared with a Fisher's exact test, continuous variables were compared with the Kruskal-Wallis test. Significance level was set at 0.05.

## Results

*Patient demographics*

Patient characteristics for the training, test and assessment datasets are shown in Table 2. No statistically significant differences were found between the groups regarding sex, age, tumor type, number of lesions, state of the non-affected liver parenchyma (i.e., steatosis) and history of previous interventions in the liver.

**Table 2:** Patient characteristics

| Characteristic | Training set | Test set | Assessment set | *P* value |
|---|---|---|---|---|
| Number of patients | 72 | 18 | 10 | |
| Sex (%) | | | | 0.534 |
|     Male | 41 (56.9) | 13 (72.2) | 6 (60.0) | |
|     Female | 31 (43.1) | 5 (27.8) | 4 (40.0) | |
| Age (years), median [range] | 61 [31 – 88] | 55 [34 – 76] | 65 [44 – 82] | 0.214 |
| Tumor type (%) | | | | 0.067 |
|     CRLM | 66 (91.7) | 14 (77.8) | 8 (80.0) | |
|     NET | 3 (4.2) | 2 (11.1) | 0 (0.0) | |
|     GIST | 2 (2.8) | 0 (0.0) | 0 (0.0) | |
|     Other | 1 (1.4) | 1 (5.6) | 2 (20.0) | |
| Number of lesions, median [range] | 5.5 [1 – 32] | 4 [1 – 14] | 7 [1 – 23] | 0.409 |
| Non-affected liver parenchyma (%) | | | | 0.559 |
|     Normal | 51 (70.8) | 12 (66.7) | 8 (80.0) | |
|     Steatosis or sinusoidal dilatation | 22 (30.6) | 6 (33.3) | 2 (20.0) | |
| Previous intervention liver (%) | | | | 0.686 |
|     None | 54 (75.0) | 14 (77.8) | 9 (90.0) | |
|     Resection and/or ablation | 18 (25.0) | 4 (22.2) | 1 (10.0) | |

*Quantitative evaluation*

Figure 2 illustrates the DSC scores for each anatomical structure. Liver parenchyma segmentation achieved the highest accuracy, with a mean DSC of 0.97±0.01. Segmentation of intrahepatic structures yielded DSC scores of 0.80±0.04 for the hepatic vein, 0.79±0.07 for the biliary tree, 0.77±0.17 for tumors, and 0.74±0.06 for the portal vein. The average tumor detection rate was 76.6±24.1% with a median of 1 false positive per patient. Vascular and biliary segmentations were categorized into central and peripheral regions. Notably, segmentation accuracy for central vascular structures was consistently higher than for peripheral vessels. Gallbladder (central biliary tree) and bile ducts (peripheral biliary tree) achieved a DSC of 0.84±0.06. Additional vasculature and tumor segmentation metrics can be found in Supplementary Material.

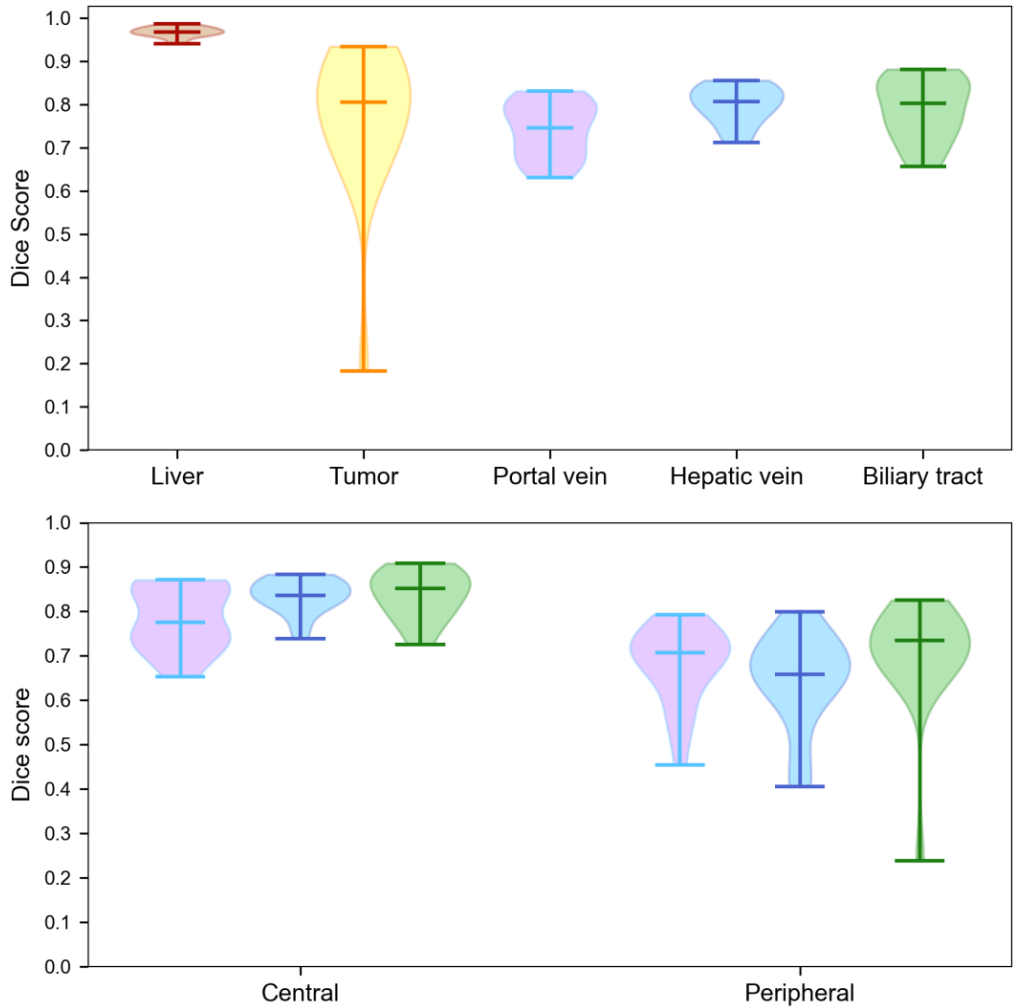

**Figure 2:** Dice similarity coefficients of the test dataset compared with manual segmentations. The violin plots illustrate the median, range, and distribution of the Dice scores. Central vascular segmentations included main trunks and primary branches directly arising from them, and peripheral regions included all smaller branches. Biliary tree segmentation included the gallbladder (central) and the bile ducts (peripheral).

Performance of the network in prospective use of the network are visualized in Figure 3, showing the extent of required manual modifications for clinical use of the 3D models (e.g., preoperative 3D planning and/or image-guided surgery). Parenchymal segmentations did not require manual refinement (DSC 1.00±0.00). Vascular structures showed DSC scores of 0.98±0.01 for the portal vein and 0.95±0.07 for the hepatic vein. The most manual adjustments were required in tumor segmentation, which demonstrated a mean DSC of 0.80±0.27. Challenging cases of tumor segmentation included patients with prior local intervention in the liver, patients with extremely high tumor burden, and low-quality scans (e.g., steatotic livers, movement artifacts). With regards to small tumors, in prospective use of the network, three sub-centimeter tumors were identified that had not been initially recognized by the radiologist (Figure 4). After consultation, a board-certified radiologist confirmed these three lesions as malignant.

The integration of automated segmentation into the clinical workflow significantly reduced the time required for manual segmentation, decreasing the overall processing time from several hours to approximately 15 minutes per patient.

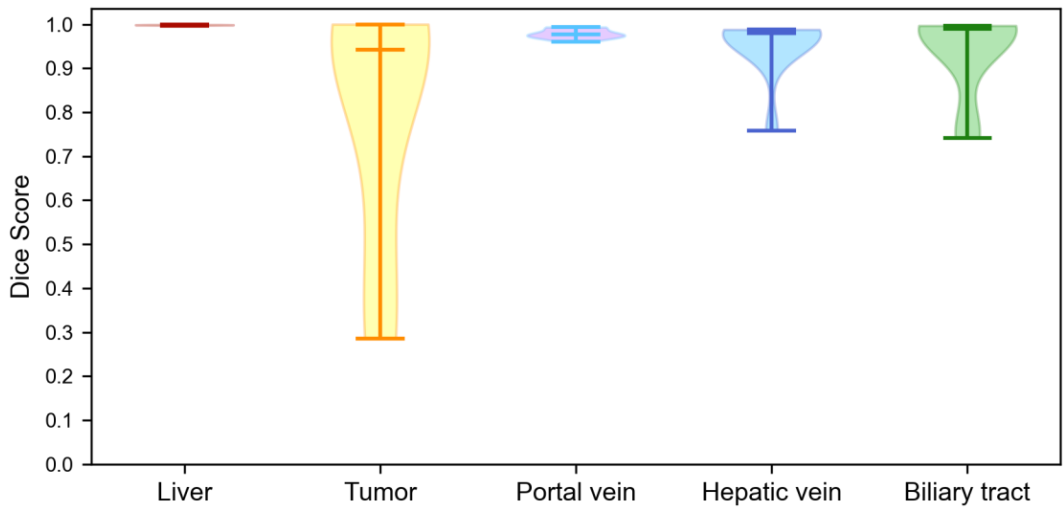

**Figure 3:** Dice similarity coefficient achieved by comparing the segmentations produced for the assessment dataset and the final manually adjusted segmentations for clinical use. Segmentations created by the model required few corrections.

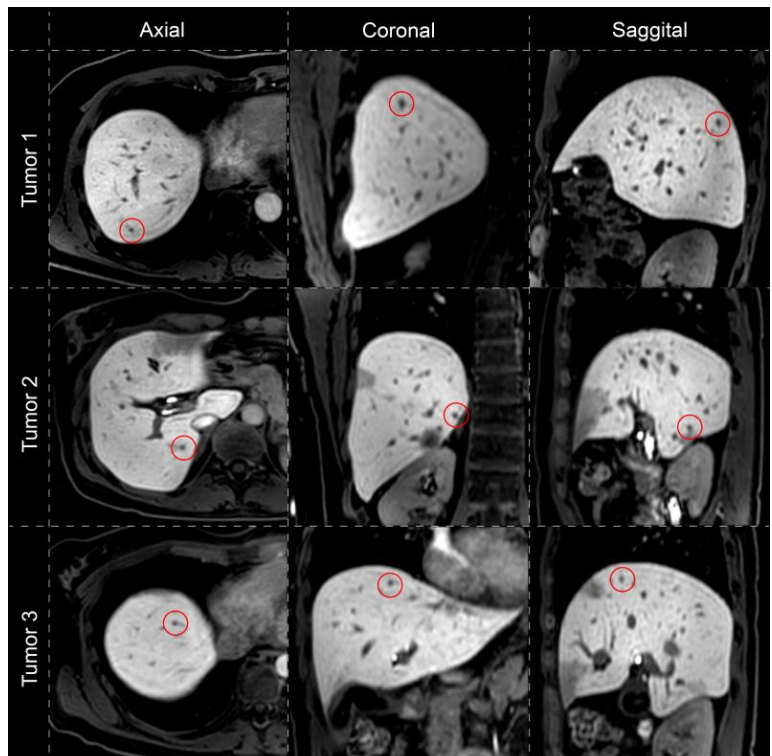

**Figure 4:** Three subcentimeter tumors initially missed by radiologists were detected by the proposed deep learning framework. Upon consultation and review, the radiologist confirmed the lesions as malignant
*Illustrative cases*

Five representative cases, with corresponding 3D models and DSC scores, are visualized in Figure 5. The highest segmentation accuracy for intrahepatic anatomy of these five cases was achieved in Patient 1. This represents an example of a high-quality MRI with a single hepatic tumor.

Patient 2 highlights a more complex scenario, characterized by severe hepatic steatosis, which significantly reduces image contrast between the liver parenchyma, vascular structures, and tumors.

The network was trained to segment malignant tumors while avoiding segmentation of benign hepatic lesions (e.g., cysts and hemangiomas), and post-treatment changes (e.g., resection sites and ablation zones). Patient 3 illustrates this task, as the patient had four hepatic cysts (indicated with green arrow) and two tumors (red arrow). The network correctly identified two cysts as benign lesions but misclassified the remaining two as tumors, highlighting both its potential and areas for further refinement in lesion characterization.

Patient 4 presents a case of extensive tumor burden, including a very large hepatic lesion with complex morphology.

A particularly challenging segmentation task is presented by patient 5, who had previously undergone a left hemi-hepatectomy. Postoperative regeneration of the organ led to neovascularization and substantial anatomical changes. Finally, the segmentation framework is capable of recognizing the absence of the gallbladder in patients who had undergone cholecystectomy. This capability was demonstrated in patients 4 and 5.

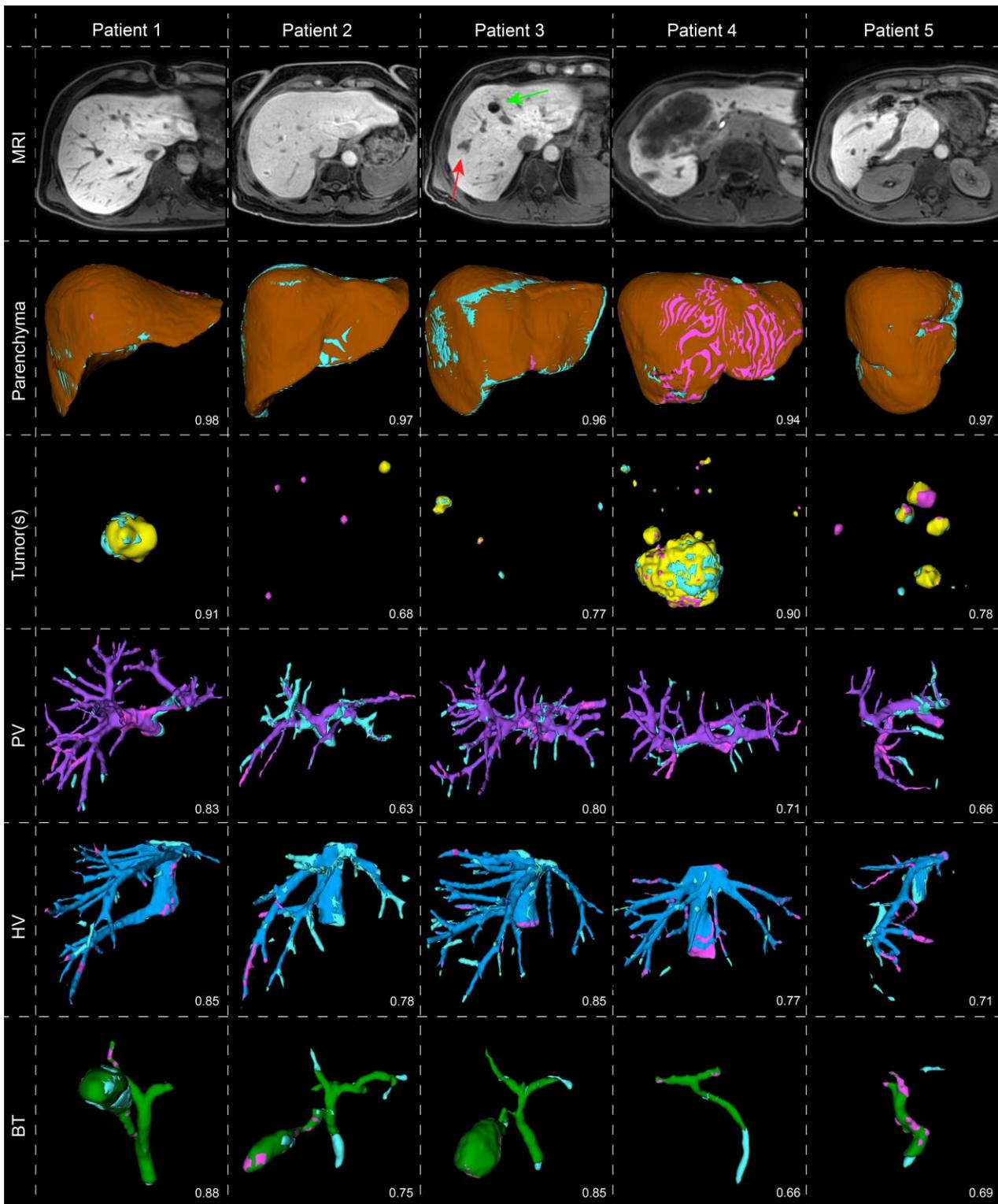

**Figure 5:** Test set results, with correct, under- (pink) and over-segmentation (blue), and corresponding DSC. Five representative cases were used to highlight the network's performance in various scenarios: a qualitatively good scan (patient 1), steatotic liver (patient 2), differentiation between tumors (red arrow) and cysts (green arrow) (patient 3), extensive tumor burden (patient 4) and altered anatomy after left hemi-hepatectomy (patient 5).

**Discussion**

This study presents a deep learning-based segmentation framework for the automatic extraction of hepatic structures from gadoxetic acid-enhanced MRI. The dataset of this study includes 3D models used for preoperative planning purposes and during image-guided liver surgery to improve intraoperative localization of small and vanished lesions. Given their critical importance in these procedures, the dataset contains detailed annotations. The proposed network demonstrated high accuracy in delineating liver parenchyma, tumors, vasculature, and bile ducts, and demonstrated robust results in a challenging patient cohort (e.g., steatotic livers, prior liver interventions, extensive tumor burden). Integration of the network into the clinical workflow effectively reduced the manual workload associated with 3D liver modelling. The high DSC scores achieved in the assessment dataset show that only minor adjustments to the automated segmentations were required for clinical implementation, shortening segmentation time from several hours to approximately 15 minutes per patient. Conventional manual segmentation methods are only available in niche centres, which have the required expertise and resources to provide such detailed anatomical representations. In contrast, this paper suggests the use of 3D planning as a standard-of-care for every patient undergoing liver surgery.

A comparison with prior studies further contextualizes these findings. Zbinden et al. (17) developed an approach for the automatic extraction of vasculature from non-contrast T1 MRI, reporting DSC scores of 0.94±0.02 for parenchyma, 0.63±0.09 for portal veins, and 0.53±0.12 for hepatic veins. Ivashchenko et al. (26) described a DVnet segmenting hepatic vasculature from contrast-enhanced MRI achieving DSC scores of 0.60±0.08 and 0.65±0.05 for portal and hepatic veins respectively. Similar to our approach, Oh et al. (18) described an automated segmentation method from hepatobiliary phase MRI, also including tumor and bile duct segmentation. They achieved DSC scores of 0.92±0.03 for parenchyma, 0.77±0.21 for tumors, 0.61±0.03 for portal veins, 0.70±0.05 for hepatic veins, and 0.58±0.15 for bile ducts. In comparison, our network achieved higher segmentation performance across all structures.

Nonetheless, automated tumor segmentation remains a critical challenge. This is due to the variable tumor appearance and size, as well as the difficulty in distinguishing tumors from benign lesions. A promising approach to enhance tumor detection would be to use the additional input of diffusion weighted MRI, as proposed by Jansen et al. (27) They achieved a tumor detection rate of 99.8%, with a median of two false positives per image, though. With the current accuracy for tumor segmentation, it is important that segmentations are always checked with annotations provided by a radiologist. While there is potential for further refinement of the proposed tumor segmentation method, its prospective application in clinical practice identified three tumors initially missed by radiologists.

There are some limitations to this study. First, no validation on an external dataset was performed. While our dataset encompasses a diverse range of complex oncological cases, external validation on a multi-centric dataset is necessary to assess the model's robustness across different populations, imaging protocols and scanner manufacturers. In addition, the current network does not include automatic segmentation of the hepatic artery and intrahepatic arterial vasculature, as it is poorly visible in the hepatobiliary phase sequence. Due to the small size of intrahepatic arteries, they are typically undetectable on intraoperative ultrasound, making them unsuitable for landmark registration and thus less critical to include in the segmentation compared to the portal and hepatic veins. For major hepatectomies, incorporating arterial segmentation may be beneficial for identifying aberrant vascular anatomy, similar to the framework proposed by Ivashchenko et al. (20,28) They developed a 4D segmentation framework for extracting hepatic anatomy from multiple phases of dynamic mDIXON MR liver.

## Conclusions

This study presents a clinically integrated deep learning model for automated segmentation of hepatic anatomy from gadoxetic acid-enhanced MRI. The network demonstrated high segmentation accuracy across parenchymal, vascular, biliary, and tumour structures, with minimal manual adjustments required for clinical application. The integration of the proposed model into preoperative planning significantly reduced segmentation time, enhancing clinical efficiency and enabling a broader access to 3D modelling beyond specialized centers. While tumour segmentation remains a key challenge and external validation is required, these results validate the integration of automated MRI-based segmentation into standard-of-care for liver surgery.

**List of abbreviations:** Dice similarity coefficient (DSC), 3-dimensional (3D), Digital Imaging and Communications in Medicine (DICOM), Neuroimaging Informatics Technology Initiative (NIfTI), portal vein (PV), hepatic vein (HV)

## Declarations

*Ethics approval and consent to participate:* This study complied with the Declaration of Helsinki and was approved by the Netherlands Cancer Institute Institutional Review Board (IRB-d24-254). Written informed consent was obtained from all subjects treated within the study protocol for surgical navigation, which was approved by the institutional medical ethics committee in July 2018 (NL65724.031.18). Patients from June 2022 and later were treated using surgical navigation as standard clinical practice at our institute based on indication.

*Consent for publication:* Not applicable.

*Availability of data and materials:* The datasets generated and/or analysed during the current study are not publicly available due institutional data sharing agreements but are available from the corresponding author on reasonable request.

*Competing interests:* The authors declare that they have no competing interests.

*Funding:* KO and MF are funded by the Koningin Wilhelmina Fonds (KWF Kankerbestrijding), Amsterdam, The Netherlands (Project ID: 12815).
BG and SS are funded by the German Research Foundation (DFG, Deutsche Forschungsgemeinschaft) as part of Germany's Excellence Strategy – EXC 2050/1 (Project ID: 390696704) – Cluster of Excellence "Centre for Tactile Internet with Human-in-the-Loop" (CeTI) of Technische Universität Dresden.

*Authors' contributions:* Conceptualization: K.O., M.F., A.Z., Data curation: K.O., K.K., A.Z., Formal analysis: K.O., B.G., A.Z. Funding acquisition: M.F., S.S., T.R. Investigation: K.O., B.G., T.N., A.Z. Methodology: K.O., A.Z. Project administration: K.O. Resources: S.S., T.R. Software: K.O., B.G., A.Z. Supervision: M.F., S.S., K.K., T.R. Validation: K.O., B.G., A.Z. Visualization: K.O., B.G., A.Z. Writing – original draft: K.O., B.G., A.Z. Writing – review and editing: K.O., M.F., B.G., T.N., B.W., S.S., T.R., K.K., A.Z.

*Acknowledgements:* Not applicable.